\documentstyle[11pt]{article}
\begin{document}
\begin{titlepage}

\vspace{0.2cm}

\title{  Probing R-parity violation in the production of
         $t \bar{c}(c \bar{t})$ on the lepton colliders
        \footnote{Supported in part by Committee of
         National Natural Science Foundation of China and Project IV.B.12 of
         scientific and technological cooperation agreement between China and
         Austria}}
\author{{ \ Yu Zeng-Hui $^{a,c}$ \ Herbert Pietschmann $^{a}$
\ Ma Wen-Gan  $^{b,c}$ \ Han Liang  $^{c}$  \ Jiang Yi  $^{c}$
 }\\\
{\small $^{a}$Institut f\" ur Theoretische Physik, Universit\" at Wien, A-1090 Vienna, Austria} \\
{\small $^{b}$CCAST (World Laboratory), P.O.Box 8730, Beijing 100080,P.R.China} \\
{\small $^{c}$Department of Modern Physics, University of Science and Technology}\\
{\small of China (USTC), Hefei, Anhui 230027, P.R.China}\\
}
\date{}
\maketitle

\vskip 12mm

\begin{center}\begin{minipage}{5in}

\begin{center} ABSTRACT\end{center}
\baselineskip 0.3in

{We studied the process $e^{+}e^{-} \rightarrow
t\bar{c}+c\bar{t}$ in a $R_{p}$-violating supersymmetric model
with effects from both B- and L-violating interactions.
The calculation shows that it is possible to either detect the 
$R_{p}$-violating
signal at the Next Linear Collider or get more stringent
constraints on the heavy-flavor $\rlap/R_{p}$ couplings.
A comparision with results from $\gamma\gamma \rightarrow t\bar{c}+
c\bar{t}$ may allow to distinguish between B- and L-violating
interactions.
For very clean background conditions and $R_{p}$ violating
parameters close to present limits, a future detection of B-violating
interactions should be possible. The process of $\mu ^{+}
\mu ^{-}
\rightarrow t \bar{c}+c\bar{t}$ is also considered.} \\

\vskip 10mm

{~~~~PACS number(s): 13.65.+i, 13.88.+e, 14.65.-q, 14.80.Dq, 14.80.Gt}
\end{minipage}
\end{center}
\end{titlepage}

\baselineskip=0.36in

\eject
\rm
\baselineskip=0.36in

\begin{flushleft} {\bf I. Introduction} \end{flushleft}
\par
In the Minimal Supersymmetric Standard Model (MSSM)\cite{s1}, R-parity
symmetry($R_{p}$) is imposed on the superfield Lagrangion to guarantee the
B- and L-conservation automatically. This symmetry is defined by
\begin{equation}
R_{p}=(-1)^{3B+L+2S}
\end{equation}
where $S$ is the spin of the particle. The discrete symmetry was
introduced\cite{s2} to avoid catastrophic proton decays from
$R_{p}$-violating interactions. In the models of R-parity
conservation, superparticles can be only pair produced and the
lightest superparticle(LSP) will be stable. Thus the LSP is a candidate
of cold dark matter in the universe.
\par
However, in order to avoid proton decays we
just need either B-conservation or L-conservation\cite{s3}. Moreover,
models of $R_{p}$ violation provide 
for neutrino masses and mixing. In those models neutrinos may get
tree-level mass contributions via mixing with gauginos and higgsinos,  
and of course also from one-loop
corrections. Unlike the general see-saw mechanism, which 
involves a high engery scale (about $10^{12}\sim 10^{16} GeV$),
we can explain neutrino masses with weak-scale physics.
With first signals for neutrino oscillations from atmospheric 
neutrinos observed
in Super-Kamiokande\cite{s4}, $\rlap /R_{p}$ is getting more and more 
interesting.
\par
Possible signals of R-parity violation in collider experiments have also
been
discussed. In the HERA $e^{+}p$ deep inelastic scattering(DIS)\cite{s5},
an anomaly has been observed. It was found that the rate of
neutral current(NC) events is higher than that predicted by the Standard
Model
when $Q^2$ is larger than $15,000 GeV^{2}$
(The possibility of a statistical fluctuation is about $10^{-3}$). For
charge
current(CC) events,
a difference between observation and prediction of SM also exists,
although
not as large as for NC events. The anomaly can be explained beautifully  
by $\rlap /R_{p}$ supersymmetric mechanism, providing a possible hint 
for R-parity violation.
\par
Because $\rlap /R_{p}$ models open many channels forbidden
or highly suppressed in $R_{p}$ conservating models, we can
get many constraints from low-energy phenomenology\cite{s6}. 
Results are collected in Ref.\cite{s7}.
\par
Let us now consider lepton colliders. Possible ways to
find a signal of $\rlap /R_{p}$ are as follows:
\par
1. Single production of sparticles and LSP decay(direct signal).
\par
2. Fermion pair productions are different in
$\rlap /R_{p}$ models and $R_{p}$ conservation \\
models(indirect signal).
\par
3.Flavor changing neutral current(FCNC) and CP violation(indirect
signal).
\par
In this paper we will concentrate on the third way. 
The process $
l^{+}l^{-}\rightarrow f_{J}f_{J^{'}}$($J$ and $J^{'}$ are different
flavors) is calculated from the L-violating terms of $\rlap /R_{p}$
models\cite{s8}.
\par
Although many constraints from low-energy phenomenology were already
given,
$\rlap /R_{p}$ parameters involving heavy flavors are not strongly
limited. With the assumption of family symmetry \cite{s9}, we can get
$\lambda_{ijk} \sim Y_{ijk}$(where $\lambda_{ijk}$ are defined
in Eq.(2.1) and $Y_{ijk}$ are
Yukawa couplings). So it
is still possible to detect them on future colliders in
the high energy region.
\par
In this paper we will use $t\bar{c}$ and $c\bar{t}$ production to
probe $\rlap /R_{p}$ signals on the Next-Linear-Collider
(NLC), First-Muon-Collider(FMC)\cite{s10} and possibly also at LEP2.
Compared
with LEP2, NLC will have much higher luminosity and energy, providing
a powerful probe. This is even more true should the FMC go into operation.
\par
Although many processes with L-violation on lepton
colliders have been calculated,
B-violation effects are rarely considered. Up to now
B-violation
parameters involving heavy flavors are still constrained weakly.
For example $\lambda^{"}_{2ij}$ and $\lambda^{"}_{3ij}$, get their
strongest
constraints from the width ratio of Z to leptons and hadrons, still being
of order one(O(1)). Hence future colliders can either detect them(if
they are close to present upper limits) or strongly improve the limits.
\par
Let us consider the possible background:
\par
1. Standard Model. 
\par
The background from SM is suppressed by the GIM mechanism.
The process of $e^{+}e^{-}\rightarrow t\bar{c}(c\bar{t})$ was considered
by C.S. Huang et al\cite{s11}. They pointed out that the cross section of
the
process is about $10^{-9}fb$ for c.m energy of about 200-500 GeV, 
thus being a negligible background for
$\rlap/R_{p}$ effects.
\par
2.Two-Higgs-Doublet-Model(THDM). 
\par
In the so called Model III of Ref.\cite{s12},
which gives the strongest effects of FCNC, the process
$e^{+}e^{+}(\mu ^{+} \mu ^{-}) \rightarrow t\bar{c}$ was considered by
Atwood et al\cite{s12},
and $\gamma\gamma \rightarrow t\bar{c}(c\bar{t}$) by Y.Jiang et
al\cite{s12}. The results show that there would be 0.1 events for 
$e^{+}e^{-}\rightarrow t\bar{c}(c\bar{t}$) and several events for   
$\gamma\gamma\rightarrow t\bar{c}(c\bar{t}$) for a luminosity
about $50 fb^{-1}$. But the effects should be much smaller,
assuming the masses of higgses to be far from 
the c.m. energy of the colliders. So
it will be easy to distinguish them from effects from
$R_{p}$ violation.
\par
3.MSSM with $R_{p}$ conservation. 
\par
Squark mixing can generate FCNC in this
model. But under the assumption of alignment of S.Dimopoulos\cite{s13}, 
it should 
be very small: mixing between up-type squarks can be as small as $10^{-3}$
to $10^{-5}$ times the KM matrix elements.
\par
In Left-Right Symmetric Models there is also a contribution to FCNC from
$Z^{'}$
decay. Because the mass of $Z^{'}$ is very large, we can omit it in our
calculations, where the c.m. engery is less than 500
GeV.   
\par
After these general remarks concerning 
the process $l^{+}l^{-} \rightarrow
t \bar{c}+c \bar{t}$,
we define the supersymmetric
$\rlap/R_{p}$ interaction in section 2. In section 3 we give the
analytical
calculations of $e^{+}e^{-} \rightarrow t \bar{c}+ c \bar{t}$.
In section 4 the numerical results of the processes $e^{+}e^{-}
\rightarrow t \bar{c}+ c \bar{t}$ and $\mu ^{+} \mu ^{-}
\rightarrow t \bar{c}+ c\bar{t}$ are presented. The conclusion
is given in section 5 and some details of the expressions are
listed in the appendix.

\par

\begin{flushleft} {\bf II. R-parity violation($\rlap/R_{p}$) in
MSSM}\end{flushleft}
\par
All renormalizable supersymmetric $\rlap/R_{p}$ interactions can
be introduced in the super potential\cite{s6}:
$$
\begin{array} {lll}
    W_{\rlap/R_{p}} & =\frac{1}{2}
\lambda_{[ij]k} L_{i}.L_{j}\bar{E}_{k}+\lambda^{'}_{ijk}
L_{i}.Q_{j}\bar{D_{k}}+\frac{1}{2}\lambda^{''}_{i[jk]}
\bar{U}_{i}\bar{D}_{j}\bar{D}_{k}+\epsilon _{i} L_{i} H_{u}.
\end{array}
\eqno {(2.1)}
$$
where $L_i$, $Q_i$ and $H_u$ are SU(2) doublets containing lepton, quark
and Higgs superfields respectively, $\bar{E}_j$ ($\bar{D}_j$, $\bar{U}_j$)
are the singlets of lepton (down-quark and up-quark),
and $i,j$ are generation indices and square brackets on them denote
antisymmetry in the bracketted indices.
\par
 We ignored the last term in Eq(2.1) because
its effects are rather small in our process\cite{s7}\cite{s14}.
So we have 9 $\lambda$-type, 27 $\lambda^{'}$-type and 9
$\lambda^{''}$-type independent parameters left. The Lagrangian density
of $\rlap/R_{p}$(to lowest order) is given as follows:
\begin{eqnarray*}
L_{\rlap/R_{p}}&=&L_{\rlap/R_{p}}^{\lambda}+L_{\rlap/R_{p}}^{\lambda^{'}}+
L_{\rlap/R_{p}}^{\lambda^{''}}
\hskip 25mm (2.2)
\end{eqnarray*}

\begin{eqnarray*}
L_{\rlap/R_{p}}^{\lambda}&=&\frac{1}{2} \lambda_{[ij]k}
[\tilde{\nu}_{iL}\bar{e}_{kR}e_{jL}+
\tilde{e}_{jL}\bar{e}_{kR}\nu_{iL}+
\tilde{e}^{*}_{kR}\bar{\nu}_{iL}^{C}e_{jL}-\\&&
\tilde{\nu}_{jL}\bar{e}_{kR}e_{iL}-\tilde{e}_{iL}\bar{e}_{kR}\nu_{jL}-
\tilde{e}^{*}_{kR}\bar{\nu}_{jL}^{C}e_{iL}]+h.c.
\end{eqnarray*}

\begin{eqnarray*}
L_{\rlap/R_{p}}^{\lambda^{'}}&=& \lambda^{'}_{ijk}[ \tilde{\nu}_{iL}\bar{d}_{kR}d_{jL}+
\tilde{d}_{jL}\bar{d}_{kL}^{C}\nu_{iL}+
\tilde{d}^{*}_{kR}\bar{\nu}_{iL}^{C}d_{jL}-\\&&
\tilde{e}_{iL}\bar{d}_{kR}u_{jL}-
\tilde{u}_{jL}\bar{d}_{kR}e_{jL}-
\tilde{d}^{*}_{kR}\bar{e}_{iL}^{C}u_{jL}]+h.c.
\end{eqnarray*}

\begin{eqnarray*}
L_{\rlap/R_{p}}^{\lambda^{"}}&=&\lambda^{"}_{i[jk]}\epsilon_{\alpha\beta\gamma}
[\tilde{u}^{*}_{iR\alpha}\bar{d}_{kR\beta}d^{C}_{jR\gamma}+
\tilde{d}^{*}_{jR\beta}\bar{u}_{iR\alpha}d^{C}_{kR\gamma}+
\tilde{d}^{*}_{kR\gamma}\bar{u}_{iR\alpha}d^{C}_{jR\beta}]+h.c.
\hskip 25mm (2.3)
\end{eqnarray*}
\par
From the interactions above, we find that only
$L_{\rlap/R_{p}}^{\lambda^{'}}$ contributes to
$l^{+}l^{-}\rightarrow t\bar{c}+c\bar{t}$ at tree-level. So the
contribution from $L_{\rlap/R_{p}}^{\lambda}$ can be neglected.
\par
The proton lifetime limit supresses the possibilities of both B-violation
and
L-violation and leads to the constraints:\cite{s7}
$$
|(\lambda~or~\lambda^{'}) \lambda^{"}|<10^{-10}(\frac{\tilde{m}}{100
GeV})^{2}.
\eqno {(2.4)}
$$
\par
The contributions from $L_{\rlap/R_{p}}^{\lambda^{"}}$ are rather weak,
but they can be separated from those of 
$L_{\rlap/R_{p}}^{\lambda^{'}}$.
Therefore, $L_{\rlap/R_{p}}^{\lambda^{"}}$ effects will 
be considered also.
\par
In the past years, many limits on the paramters $\lambda$,
$\lambda^{'}$ and $\lambda^{"}$ were given from low-energy experiments.
The upper limits were calculated with the assumption that only one
coupling parameter is non-zero\cite{s15}. On that basis, the parameters
$\lambda$, $\lambda^{'}$ and $\lambda^{"}$ are typically less than
$10^{-1}-10^{-2} (\frac{\tilde{m}}{100 GeV})^{2}$\cite{s7}. Although
some authors argue that the limits can be relaxed\cite{s16} if 
the so-called single coupling hypothesis is dropped, we shall use these
upper bounds in our paper.

\begin{flushleft} {\bf III. Calculations } \end{flushleft}
\par
In the following calculations we assume the parameters $\lambda^{'}$
and $\lambda^{"}$ to be real. We will only consider the lowest order 
effects from $L_{\rlap/R_{p}}^{\lambda^{'}}$ and
$L_{\rlap/R_{p}}^{\lambda^{"}}$.
\par
A. $e^{+}(p_3)e^{-}(p_4)\rightarrow t(p_1) \bar{c}(p_2)$ at tree-level.
\par
We define the Mandelstam variables as usual
$$
    s  = (p_{1}+p_{2})^2=(p_{3}+p_{4})^2
\eqno {(3.a.1)}
$$
$$
    t  = (p_{1}-p_{3})^2=(p_{4}-p_{2})^2
\eqno {(3.a.2)}
$$
$$
    u  = (p_{1}-p_{4})^2=(p_{3}-p_{2})^2
\eqno {(3.a.3)}
$$
\par
The amplitude(as shown in Fig.1.a) is given by:
\begin{eqnarray*}
M& = & \Sigma _{j}\frac{i \lambda^{'}_{13j}\lambda^{'}_{12j}}
   {(t-m_{squark_{j}}^{2})}
   \bar{u}(p_1) P_{R} u^{c}(p_3) \bar{v^{c}}(p_4) P_{L} v(p_2).
~~~~~~~~~~~~~~~~~(3.a.4)
\end{eqnarray*}
where $P_{L,R}$ are left- and right-helicity projections respectively,
$j=1,2,3$ and the upper index $c$ means charge conjugate. The
amplitude
depends strongly on the products
$\lambda^{'}_{12j}\lambda^{'}_{13j}$($j=1,2,3$).
\par
B. Contributions from $L_{\rlap/R_{p}}^{\lambda^{"}}$ terms.
\par
If we set all $\lambda^{'}$ parameters to zero, we obtain
the effects of $L_{\rlap/R_{p}}^{\lambda^{"}}$ terms
within the present upper bounds.
One-loop corrections (as shown in Fig.1.b)
of $e^{+}(p_3)e^{-}(p_4) \rightarrow t(p_{1})\bar{c}(p_{2})$
are proportional to the products
$\lambda^{"}_{2ij}\lambda^{"}_{3ij}$($i,j=1,2,3$), thus it is possible
to detect $\rlap/R_{p}$ signals or get much stronger constraints on those
parameters by measuring this process in future experiments.
\par
Since the proper vertex counterterm should cancel with the counterterms
of the external legs diagrams in this case, we do not have to deal with
the ultraviolet divergence. Thus we simply take the sum of all
(unrenormalized)
reducible and irreducible diagrams and the result is finite and gauge
invariant. In the Appendix we will give the details of the amplitudes.
\par
C. Total cross sections
\par
In a similar way we obtain the amplitude for process $e^{+}e^{-}
\rightarrow c \bar{t}$. Thus the total cross section for the
process $e^{+}e^{-} \rightarrow t \bar{c}+ c \bar{t}$ is:
\begin{eqnarray*}
\hat{\sigma}(\hat{s}) = \frac{2N_{c}}{16 \pi \hat{s}^2 }
             \int_{\hat{t}^{-}}^{\hat{t}^{+}} d\hat{t} {\bar{\sum}_{spins}^{}}
             [|M|^{2}],
~~~~~~~~~~~~~~~~~~~~~~~~~~~(3.5)
\end{eqnarray*}
where $M$ is the amplitude and $\hat{t}^{\pm}=\frac{1}{2}\left[
(m_t^2+m_c^2-\hat{s})\pm \sqrt{\hat{s} ^2+m_t^4+m_c^4-2\hat{s} m_t^2-2\hat{s}
m_c^2-2 m_t^{2} m_c^{2}} \right]$. Here we have neglected the masses of
electron and
muon. $N_{c}=3$ is the color factor and the bar over summation means
averaging over initial spins.
\par
Similarly we obtain the total cross section of $\mu^{+} \mu^{-}
\rightarrow t\bar{c}+c\bar{t}$. Assuming values for all input parameters,
we obtain numerical results.

\begin{flushleft} {\bf IV. Numerical results} \end{flushleft}
\par
In the numerical calculations we assume $m_{\tilde{q}}=
m_{\tilde{l}}$ and consider the effects from
$L_{\rlap/R_{p}}^{\lambda^{'}}$
and $L_{\rlap/R_{p}}^{\lambda^{"}}$ separately.
For the B-violating parameter
$\lambda^{"}_{2ij}\lambda^{"}_{3ij}$($i,j=1-3$),
the upper bounds of $\lambda^{"}_{223}$ and $\lambda^{"}_{323}$ dominate
all other parameters. Thus we neglect all other $\lambda^{"}$
terms. For the L-violating parameters we set
$\lambda^{'}_{12j}=\lambda^{'}_{13j}=0.1$($j=1,2,3$)
when $m_{\tilde{q}}=100~GeV$,
which agrees with the product coupling limits also.
For the $\mu^{+}\mu^{-}$ colliders, 
the parameters $\lambda^{'}_{22j}$ and $\lambda^{'}_{23j}$ can be larger
because they involve heavier flavor.
In this case we use the data of reference\cite{s6}.
\par
In Fig.2, we show the cross section of $e^{+}e^{-}\rightarrow
t\bar{c}+c\bar{t}$ as function of c.m.
energy of the electron-positron system at the upper bounds of
$\lambda^{'}$,
i.e. $\lambda^{'}_{12j}\lambda^{'}_{13j}=0.01$. We take
$m_{\tilde{l}}=m_{\tilde{q}}=100~GeV$ (solid
line) and $m_{\tilde{l}}=m_{\tilde{q}}=150~GeV$ (dashed line),
respectively. There we take same coupling parameters for different
$m_{\tilde{q}}$ for comparing the effects of mass of squarks in the process.
The results show that the cross sections can be $0.02~pb$ for solid line and
$0.006~pb$ for dashed line at $\sqrt{s}=190 GeV$, which
is the present LEP running energy. So if the
electron-positron integrated luminosity is $150 pb^{-1}$\cite{s7}, we can
expect about 3 events when $m_{\tilde{l}}=m_{\tilde{q}}=100~GeV$.
At $\sqrt{s}=200~GeV$ and luminosity about $200~
pb^{-1}$, we expect 8 events from our results.
Even if this sounds too optimistic, it 
may be worthwhile to consider this process once the LEP energy is 
above the threshold of single top-quark production.
For the NLC, with c.m.energy about $500~GeV$ and luminosity about
$50~fb^{-1}$, thousands of events should be observed at the present upper
bounds of the parameters.
\par
In Fig.3, we plot the cross section of $\mu ^{+} \mu ^{-} \rightarrow
t\bar{c}+c\bar{t}$ as function of c.m. energy of the $\mu ^{+} \mu ^{-}$
system with the upper bounds of $\lambda^{'}$, i.e. $\lambda^{'}_{22j}=0.18$
and $\lambda^{'}_{23j}=0.36$ (see Ref.\cite{s6}).
We take again $m_{\tilde{l}}=m_{\tilde{q}}=100~GeV$ for the solid line and
$m_{\tilde{l}}=m_{\tilde{q}}=150~GeV$ for the dashed line.
The cross sections are much larger than those of Fig.2.
That is because from present data the upper limits of
$\lambda^{'}_{22j}$ and $\lambda^{'}_{23j}$ are larger than those of
$\lambda^{'}_{12j}$ and $\lambda^{'}_{13j}$.
The cross section can be about $1~pb$ when $\sqrt{s}=200~GeV$,
which means we can get hundreds of events at $\mu$ colliders with
the same luminosity as LEP, if the coupling parameters are close to
present upper limits.
\par
In order to give more stringent constraints for $\lambda^{"}$ in
future experiments, we draw
the effects from possible B-violating terms in Fig.4, where
the cross section of $e^{+}e^{-}\rightarrow t\bar{c}+c\bar{t}$ as
function of c.m.energy is given.(The solid line is for $m_{\tilde{l}}=
m_{\tilde{q}}=100~GeV$ and dashed line for $m_{\tilde{l}}=
m_{\tilde{q}}=150~GeV$). When $\lambda^{"}_{223}\lambda^{"}_{323}$ is
about 0.625 (see Ref.\cite{s6}), the cross section will be about
$0.5~fb$ at $\sqrt{s}=200~GeV$ or $0.9fb$ at $\sqrt{s}=500~GeV$. That
corresponds to 0.1 event at LEP or 45 events at the NLC.
\par
Let us compare
the results with those from $\gamma\gamma \rightarrow
t\bar{c}+c\bar{t}$ of Ref.\cite{s17}.
It turns out that B-violating terms (i.e.
$L_{\rlap/R_{p}}^{\lambda^{"}}$)
give similar effects in both processes, whereas L-violation(i.e.
$L_{\rlap/R_{p}}^{\lambda^{'}}$) contributes much less in $\gamma\gamma$
collisions than in $e^{+}e^{-}$ processes. Therefore, a combination
of the results of both these processes allows for a determination of the
source for $R_{p}$-violation (i.e. either from L-violation or from
B-violation)

\begin{flushleft} {\bf IV. Conclusion} \end{flushleft}
\par
We studied the processes $e^{+}e^{-} \rightarrow
t\bar{c}+c\bar{t}$
and $\mu^{+}\mu^{-} \rightarrow t \bar{c}+c\bar{t}$
in a supersymmetric model with explicit $R_{p}$-violation.
The calculations show that it is possible to test
the model at future LEP and the
future NLC experiments, provided the couplings($\lambda^{'}$-type)
are large enough within the present experimentally admitted range.
We can even detect possible B-violating terms in future lepton
colliders with higher energy and higher luminosity than LEP.
We also considered the possibility of production of $t\bar{c}$ and
$c\bar{t}$ at $\mu^{+}\mu^{-}$ colliders. The results show that these
colliders may allow to test $R_{p}$
violation.
\par
The authors would like to thank Prof. H.Stremnitzer for reading the
manuscript.
\newpage
\begin{center} {\bf Appendix} \end{center}

\par
A. Loop integrals:
\par
We adopt the definitions of two- and three- one-loop
Passarino-Veltman integral functions in reference\cite{s18}\cite{s19}.
The integral functions are defined as follows:
\par
1. The two-point integrals are:
$$
\{B_0;B_{\mu};B_{\mu\nu}\}(p,m_1,m_2)=
{\frac{(2\pi\mu)^{4-n}}{i\pi^2}}\int d^n q
{\frac{\{1;q_{\mu};q_{\mu}q_{\nu}\}}{[q^2-m_1^2][(q+p)^2-m_2^2]}},
~~~~~(A.a.1)
$$
The function $B_{\mu}$ should be proportional to $p_{\mu}$:
$$
B_{\mu}(p,m_{1},m_2)=p_{\mu} B_{1}(p,m_1,m_2)
~~~~~(A.a.2)
$$
Similarly we get:
$$
B_{\mu\nu}=p_{\mu}p_{\nu} B_{21}+g_{\mu\nu} B_{22}
~~~~~(A.a.3)
$$
We denote $\bar{B}_{0}= B_{0}-\Delta$, $\bar{B}_{1}= B_{1}+\frac{1}{2}\Delta$
and $\bar{B}_{21}= B_{21}-\frac{1}{3}\Delta$. with $\Delta= \frac{2}{\epsilon}
-\gamma +\log (4\pi)$, $\epsilon=4-n$. ${\mu}$ is the scale parameter.
\par
2. Three-point integrals:
$$
\{C_0;C_{\mu};C_{\mu\nu};C_{\mu\nu\rho}\}(p,k,m_1,m_2,m_3)=
$$
$$
-{\frac{(2\pi\mu)^{4-n}}{i\pi^2}}\int d^n q
{\frac{\{1;q_{\mu};q_{\mu}q_{\nu};q_{\mu}q_{\nu}q_{\rho}\}}
{[q^2-m_1^2][(q+p)^2-m_2^2][(q+p+k)^2-m_3^2]}},
~~~~~(A.a.4)
$$
We can express the tensor integrals through scalar functions in the 
following way:
$$
C_{\mu}=p_{\mu} C_{11} + k_{\mu} C_{12}
$$
$$
C_{\mu\nu}=p_{\mu} p_{\nu} C_{21}+k_{\mu}k_{\nu} C_{22}+
(p_{\mu}k_{\nu}+k_{\mu}p_{\mu}) C_{23}+ g_{\mu\nu} C_{24}
$$
$$
C_{\mu\nu\rho}=p_{\mu}p_{\nu}p_{\rho} C_{31}+
               k_{\mu}k_{\nu}k_{\rho} C_{32}+
               (k_{\mu}p_{\nu}p_{\rho} +
                p_{\mu}k_{\nu}p_{\rho} +
                p_{\mu}p_{\nu}k_{\rho}) C_{33}+
$$
$$
               (k_{\mu}k_{\nu}p_{\rho} +
                p_{\mu}k_{\nu}k_{\rho} +
                k_{\mu}p_{\nu}k_{\rho}) C_{34}+
                (p_{\mu} g_{\nu\rho}+p_{\nu} g_{\mu\rho}+
               p_{\rho} g_{\mu\nu}) C_{35}+
$$
$$
                (k_{\mu} g_{\nu\rho}+k_{\nu} g_{\mu\rho}+
                k_{\rho} g_{\mu\nu}) C_{36}
~~~~~~~~~~~~~~~(A.a.5)
$$

\par
The numerical calculation of the vector and tensor loop integral functions
can be traced back to the four scalar loop integrals $A_0$, $B_0$ and $C_0$
in Ref.\cite{s12}\cite{s13} and the references therein.
\par
B. one-loop correction of the amplitude.
\par
The amplitude of one-loop diagrams $\delta M$ from
$L_{\rlap/R_{p}}^{\lambda ^{''}}$
(Fig.1.b) can be decomposed into $\delta M_{\gamma}$ and
$\delta M_{Z}$ terms with:

\begin{eqnarray*}
\delta M_{\gamma} & = & \frac{e g_{\mu\nu}}
   {s}  \bar{v} (p_3) \gamma ^{\nu} u(p_4)
  \bar{u}(p_1)\Sigma^{\nu}_{\gamma}(p_1,p_2)v(p_2)
~~~~~~~~~~~~~~~~~(A.b.1)
\end{eqnarray*}
and
\begin{eqnarray*}
\delta M_{Z} & = &(\frac{e}{4c_{w}s_{w}}) \frac{g_{\mu\nu}-
k_{\mu} k_{\nu}/m_{z}^2}{s-m_{Z}^2}\\&&
    \bar{v} (p_3) \gamma ^{\nu}((2-4s_{w}^2)P_{L}-4s_{w}^2 P_{R}) u(p_4)
  \bar{u}(p_1)\Sigma^{\nu}_{Z}(p_1,p_2)v(p_2)
~~~~~~~~~~~~~~~~~(A.b.2)
\end{eqnarray*}

where $k=p_1+p_2$, $\frac{e^2}{4\pi}=\alpha=1/137.04$, $c_{w}=\cos
\theta_{W}$,
$s_{w}=\sin \theta_{W}$ and $\theta_{W}$ is the Weinberg-angle and
$\Sigma^{\nu}_{\gamma,Z}(p_1,p_2)$ is defined as follows:
\begin{eqnarray*}
\Sigma ^{\nu}_{\gamma,Z}(p_1,p_2) ~=~
V_{\gamma,Z}^{(1)} P_{R}\gamma^{\nu}+
V_{\gamma,Z}^{(2)} P_{R}p_{1}^{\nu}+
V_{\gamma,Z}^{(3)} P_{R}p_{2}^{\nu}+\\
V_{\gamma,Z}^{(4)} P_{L}\gamma^{\nu}+
V_{\gamma,Z}^{(5)} P_{L}p_{1}^{\nu}+
V_{\gamma,Z}^{(6)} P_{L}p_{2}^{\nu}
~~~~~~~~~~~~~~~~~~~~ {(A.b.3)}
\end{eqnarray*}

Here the $V_{\gamma,Z}^{(i)}$ are scalar functions of $p_1,p_2$.
\par
%\newpage

\newpage

\begin{center}
{\large \bf Figure Captions}
\end{center}

\parindent=0pt
\par
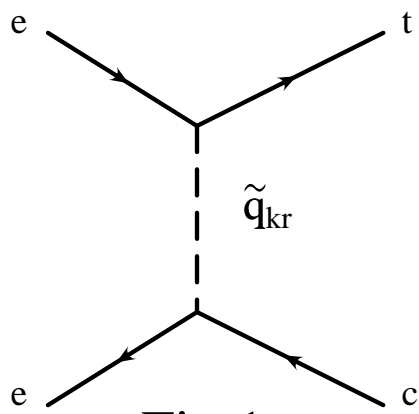
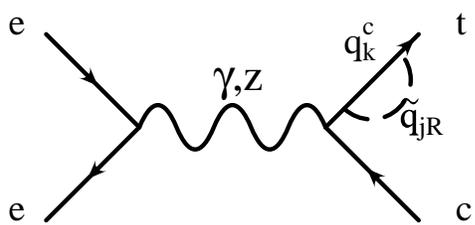
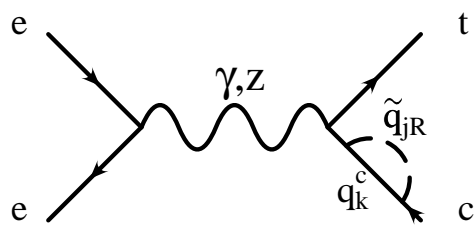
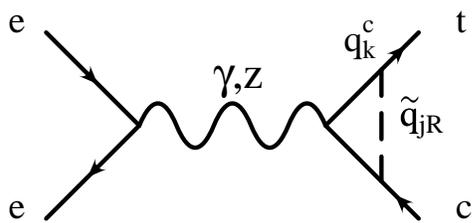
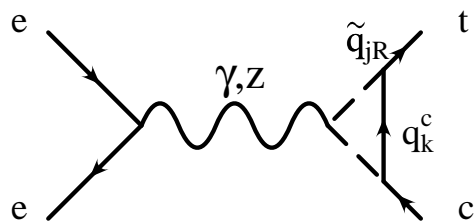
{\bf Fig.1} Feynman diagrams of $e^{+}e^{-}\rightarrow t \bar{c}$
        Fig.1 a: Tree-level diagrams from $L_{\rlap/R_{p}}^{\lambda^{'}}$.
        Fig.1 b: one-loop diagrams from $L_{\rlap/R_{p}}^{\lambda^{''}}$,
        dashed lines represent sleptons and squarks.
\par
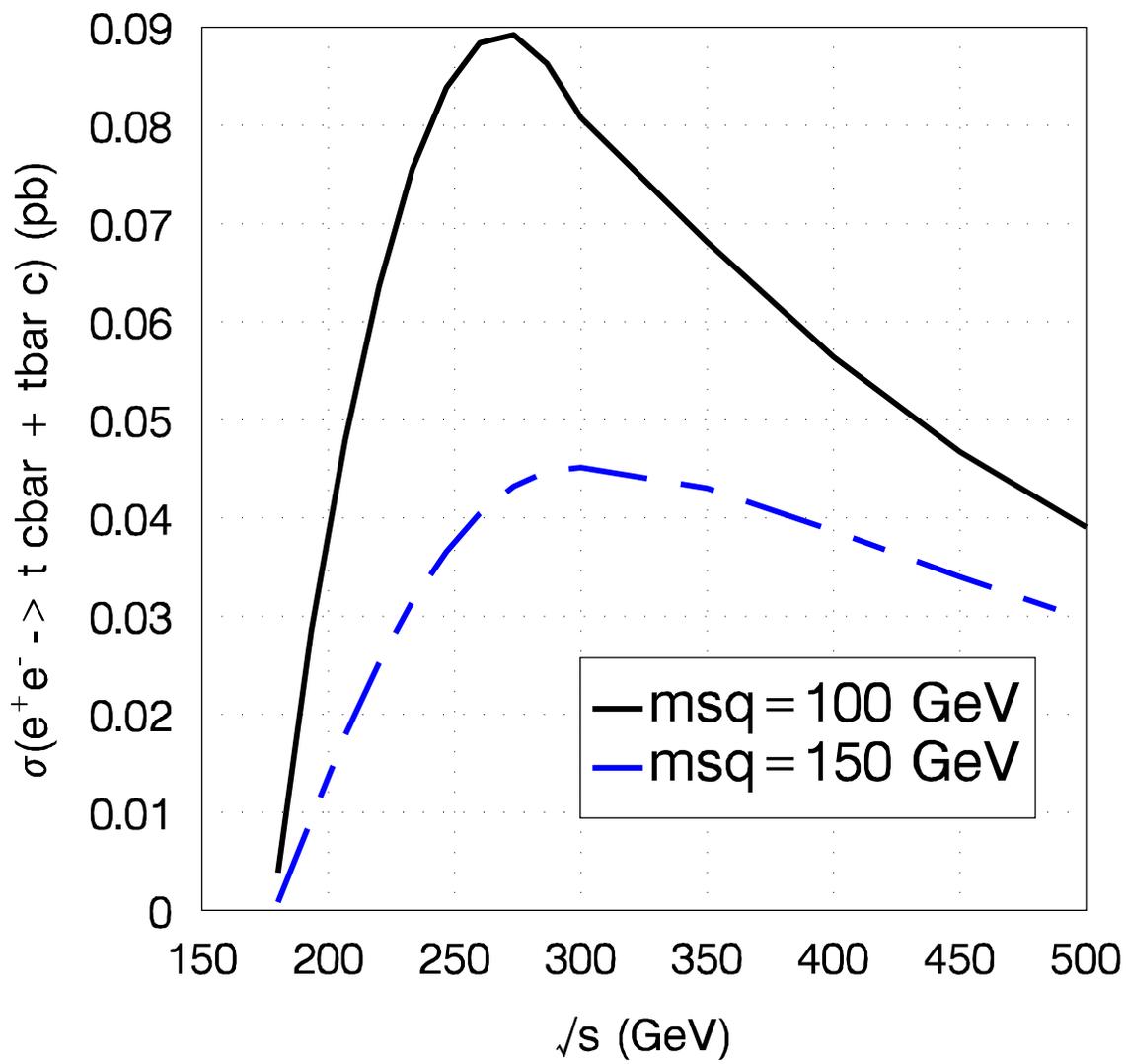
{\bf Fig.2} Cross section of $e^{+}e^{-}
        \rightarrow t\bar{c}+c\bar{t}$ as function of c.m.energy
        $\sqrt{s}$
        with $\lambda^{'}_{12j}\lambda^{'}_{13j}=0.01$
        solid line for $m_{\tilde{l}}=m_{\tilde{q}}=100~GeV$,
        and dashed line for $m_{\tilde{l}}=m_{\tilde{q}}=150~GeV$.
\par
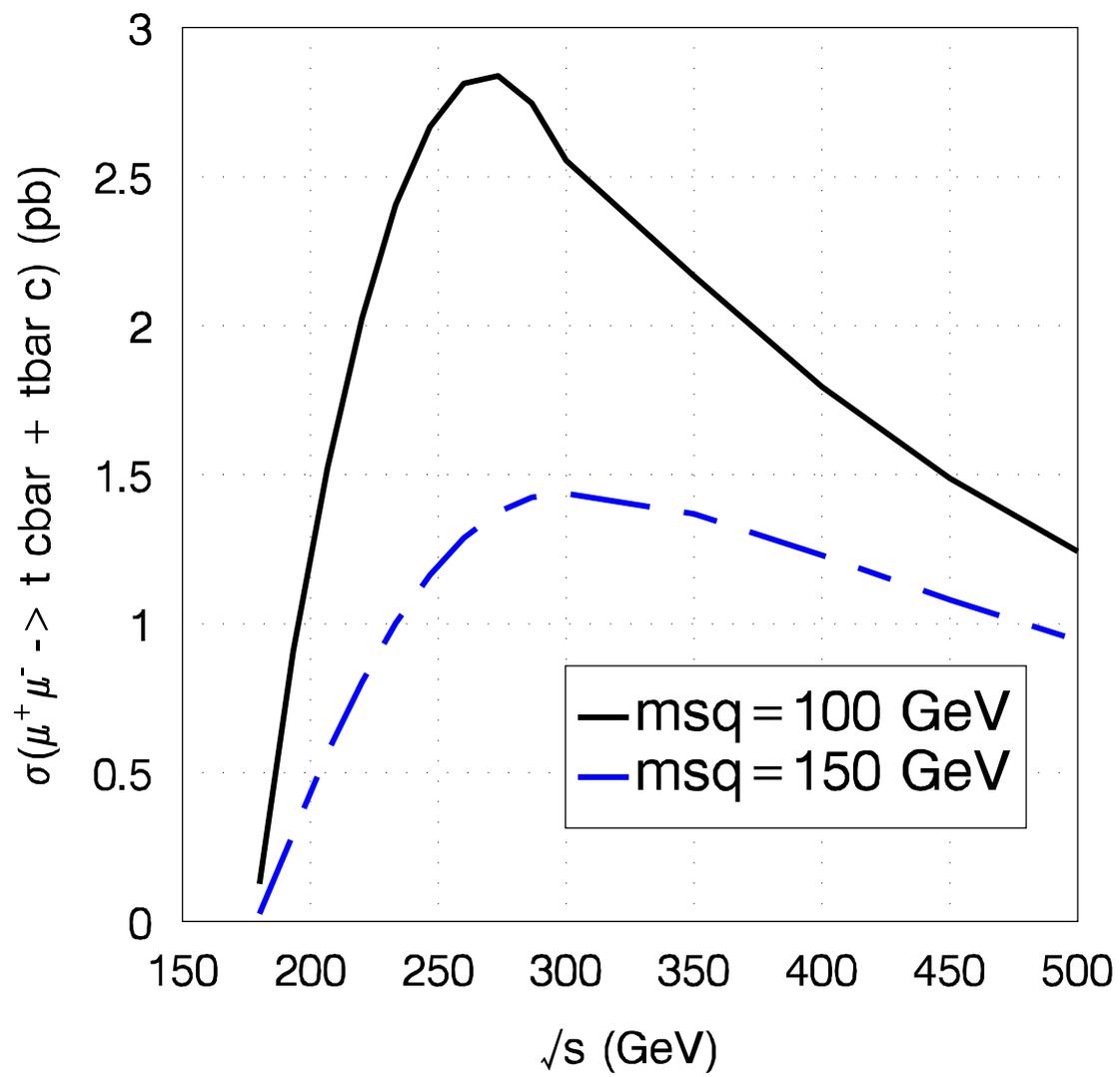
{\bf Fig.3} Cross section of $\mu ^{+}\mu ^{-}
        \rightarrow t\bar{c}+c\bar{t}$ as function of c.m.energy
        $\sqrt{s}$
        with $\lambda^{'}_{22j}=0.18$ and $\lambda^{'}_{23j}=0.36$, see
        Ref.\cite{s5}.
\par
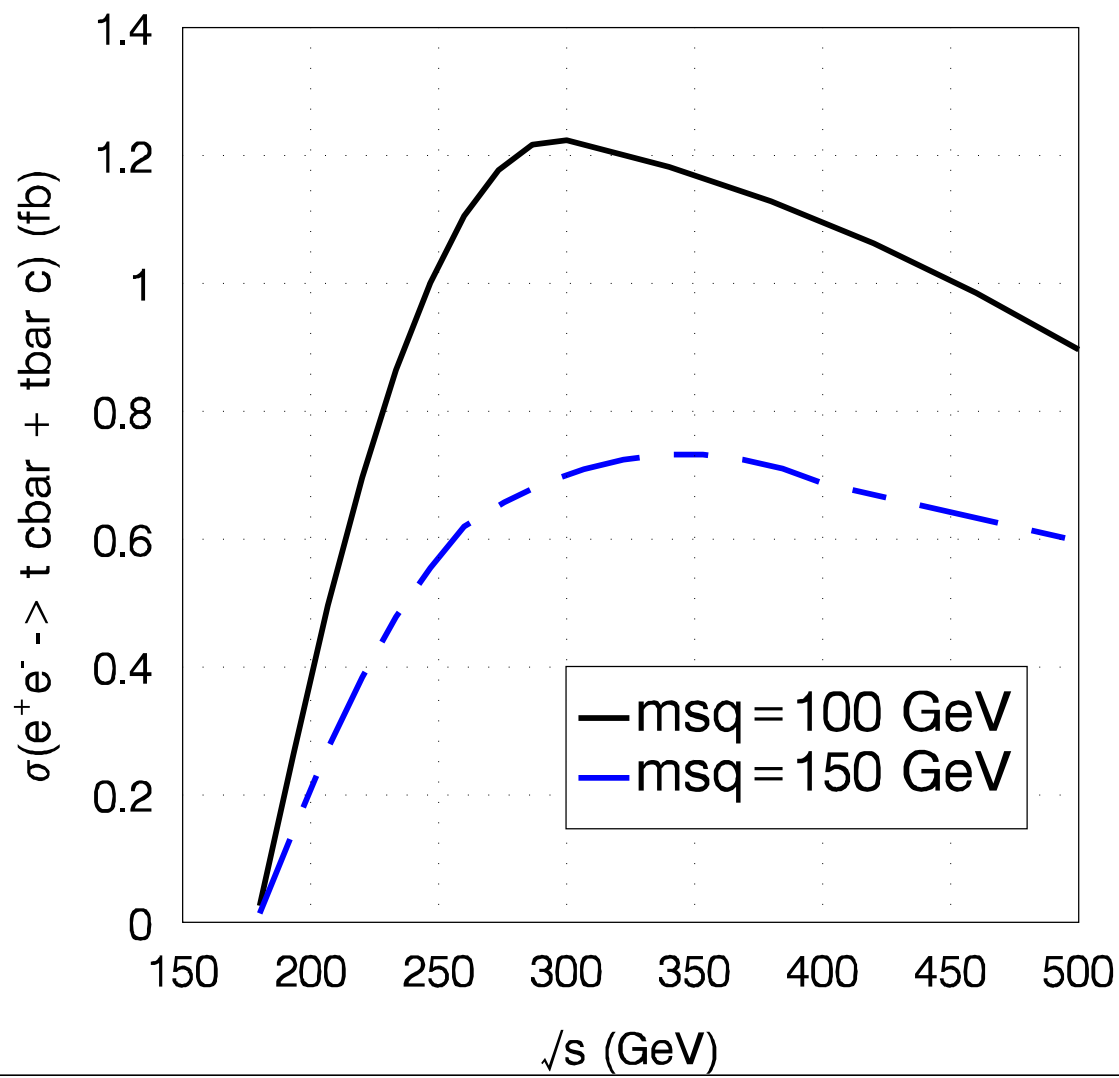
{\bf Fig.4} Cross section of $e^{+}e^{-}
        \rightarrow t\bar{c}+c\bar{t}$ as function of c.m.energy $\sqrt{s}$
        with $\lambda^{''}_{323}\lambda^{''}_{223}=0.625$
        solid line for $m_{\tilde{l}}=m_{\tilde{q}}=100~GeV$,
        and dashed line for $m_{\tilde{l}}=m_{\tilde{q}}=150~GeV$.
\end{document}